\documentclass[aps,twocolumn,superscriptaddress]{revtex4-1}

\usepackage{amssymb}
\usepackage{amsthm}
\usepackage{physics}
\usepackage{bm}
\usepackage{graphicx}

\newtheorem{lemma}{Lemma}

\begin{document}
\preprint{}

\title{Necessary Condition for Steerability of Arbitrary Two-Qubit States with Loss}

\author{Travis J. Baker}
\affiliation{Centre for Quantum Dynamics, Griffith University, Brisbane 4111, Australia}
\affiliation{Centre for Quantum Computation and Communication Technology, Griffith University, Brisbane 4111, Australia}
\author{Sabine Wollmann}
\affiliation{Centre for Quantum Dynamics, Griffith University, Brisbane 4111, Australia}
\affiliation{Centre for Quantum Computation and Communication Technology, Griffith University, Brisbane 4111, Australia}
\affiliation{Quantum Engineering Technology Labs, H. H. Wills Physics Laboratory and Department of Electrical \& Electronic Engineering, University of Bristol, BS8 1FD, UK}
\author{Geoff J. Pryde}
\affiliation{Centre for Quantum Dynamics, Griffith University, Brisbane 4111, Australia}
\affiliation{Centre for Quantum Computation and Communication Technology, Griffith University, Brisbane 4111, Australia}
\author{Howard M. Wiseman}
\affiliation{Centre for Quantum Dynamics, Griffith University, Brisbane 4111, Australia}
\affiliation{Centre for Quantum Computation and Communication Technology, Griffith University, Brisbane 4111, Australia}

\date{\today}

\begin{abstract}
Einstein-Podolsky-Rosen steering refers to the quantum phenomenon whereby the state of a system held by one party can be ``steered'' into different states at the will of another, distant, party by performing different local measurements.
Although steering has been demonstrated in a number of experiments involving qubits, the question of which two-qubit states are steerable remains an open theoretical problem.
Here, we derive a necessary condition for any two-qubit state to be steerable when the steering party suffers from a given probability of qubit loss.
Our main result finds application in one-way steering demonstrations that rely upon loss.
Specifically, we apply it to a recent experiment on one-way steering with projective measurements and POVMs, reported by Wollmann {\em et. al.} [{\em Phys. Rev. Lett.}, {\bf 116}, 160403 (2016)].
\end{abstract}


\maketitle

\section{Introduction\label{sec:intro}}

Quantum steering is a remarkable feature of quantum mechanics first noted by Einstein, Podolsky and Rosen \cite{EPR35}, and Schr\"odinger \cite{Sch35} in 1935, whereby one party (Alice) can influence the outcomes of a distant party (Bob) by performing local measurements on a shared state. 
More recently, it was shown by one of us and coworkers \cite{Wis07} that the phenomenon of steering is strictly intermediate between entanglement and Bell nonlocality---Alice's outcomes are allowed to be determined by local random variables, while it is assumed that Bob's measurements and outcomes are described by quantum mechanics.

To date, a large portion of theoretical papers on the topic of steering have focussed on the construction of steering inequalities.
These are conditions which, when violated, are sufficient to show that Alice has steered Bob.
However, the converse question can also be asked: how can it be shown that a state is {\em non-steerable}?
By definition, such a proof requires the construction of a so-called local-hidden-state (LHS) model \cite{Wis07} for the party being steered.
That is, a state is non-steerable if the correlations between Alice's (untrusted) and Bob's (trusted) observations can be simulated by some local cheating strategies and measurements on LHSs, respectively.
This question is crucial to understanding the phenomena of one-way steerability, in which Alice cannot steer Bob, yet Bob can steer Alice for a given state.

One-way steerability (1WS) has attracted considerable attention recently \cite{Han12,Bow14,Eva14a,Wol16,Xia17}.
Although the existence of Gaussian quantum states exhibiting 1WS was first shown in 2012, this was under the restriction that Alice could only make Gaussian measurements \cite{Han12}.
Two-qubit states which exhibit 1WS while allowing Alice to make arbitrary projective measurements were shown by example in 2014 \cite{Bow14}.
In the same year, it was theoretically pointed out \cite{Eva14a} that, in the context of loss-depleted two-qubit states, 1WS can be achieved more simply by passing the untrusted party's system through a sufficiently lossy channel.
The {\em amount} of qubit loss---an inevitable reality for experiments involving photons---was a central issue in finding parameter regions for 1WS.
In 2016, three of us  and co-workers reported an experiment adopting this strategy \cite{Wol16}, where two-qubit Werner states were passed through lossy channels.
The experiment produced states that were well-approximated by Werner states under a fidelity measure, and assumed, in the analysis, that the latter were the true states.
A fully rigorous demonstration of 1WS, however, requires careful consideration of deviations from the ideal correlations.
This is the motivation behind the present work.

In this paper, we construct LHS models for loss-depleted two-qubit states.
The remainder is organised as follows.
In \S \ref{assemblages} we describe the steering scenario between two parties.
In \S \ref{lossyw}, we modify this scenario to allow consideration of two-qubit states passing through a loss channel, before constructing a LHS model for this class in \S \ref{lhsmodel}.
We extend our sufficient condition for projective measurements to the most general kind of measurements (POVMs) in \S \ref{povms}, and apply it to the 1WS of experimental states reported in Ref. \cite{Wol16}.
Lastly, we draw conclusions.

\section{EPR-steering\label{assemblages}}

The steering scenario in which we are interested involves two spatially distant parties, Alice and Bob.
Let $\rho \in \mathfrak{B}\left( \mathcal{H}_A \otimes \mathcal{H}_B \right)$ be a bipartite quantum state shared between them. 
Suppose Alice can perform a collection of measurements $\{ \mathcal{M}_{\bm{x}} : {\bm{x}} \}$ on her system, where the individual measurements are indexed by $\bm{x}$.
In general, such a measurement is described by a set of positive operators $\{ E_{X | \bm{x}} : X \}$, where $E>0, \sum_X E_{X | \bm{x}} = I$, and $X$ labels the outcomes.
That is, $E_{X| \bm{x}}$ is the effect associated with a Positive Operator-Valued Measure (POVM).
Bob requests that Alice make a particular measurement on her system, and announce the result.
He then performs appropriate measurements to obtain information about his state.
Provided his set of measurements is tomographically complete, the state can be reconstructed over many repetitions of the protocol.
The set of Bob's reduced states, together with Alice's announced outcomes given the requested measurement is called an {\em assemblage}, which we denote by $\{ \{ \sigma_{X | \bm{x}} : X \} : \bm{x} \}$.
It comprises states
\begin{equation}
	\sigma_{X | \bm{x}} = \Tr_A \left[ E_{X| \bm{x}} \otimes I \rho \right], \label{eq:2assemblage}
\end{equation}
which are sub-normalized in the sense that $\Tr [ \sigma_{X | \bm{x}} ] = p(X| \bm{x})$.

The question of EPR-steerability concerns the types of assemblages for which Bob will be {\em convinced} that Alice is steering his state.
That is, Bob---without making assumptions as to how Alice generated her results---must decide whether his assemblage can be described by a LHS model \cite{Cav09}.
Formally, the shared state $\rho$ is said to be non-steerable, if and only if, for all allowed measurements $\{ M_{X | \bm{x}} : X\}$, there exists a LHS model for Bob, {\em i.e.} whether there exists an ensemble $\{ \sigma_\lambda : \lambda \}$ and positive distribution $p(X|\bm{x},\lambda)$ such that $ \forall X, \bm{x}$,
\begin{equation}
	\sigma_{X|\bm{x}} = \sigma_{X| \bm{x}}^\text{LHS} \equiv \int d\lambda p(X| \bm{x},\lambda) \sigma_\lambda. \label{eq:lhsmodel}
\end{equation}
Here, the set of Bob's LHSs $\{ \sigma_\lambda : \lambda \}$ is indexed by hidden variables $\lambda$ (known to Alice) and satisfies $\int d\lambda \sigma_\lambda =\rho_B$.
If such a model exists, the assemblage could have arisen from a dishonest Alice, attempting to fool Bob using her knowledge of Bob's LHSs, and employing a cheating strategy by which she announces outcome X given input $\bm{x}$.
On the contrary, the absence of a LHS model implies Alice will be able to convince Bob that she has steered his state.

\section{Loss-Depleted Two-Qubit States \label{lossyw}}

\subsection{Two-Qubit States}

An arbitrary two-qubit state $\rho_0$ can be expressed in terms of the Pauli matrices as
\begin{equation} 
	\rho_0 = \frac{1}{4}\left( I \otimes I + \bm{a} \cdot \bm{\sigma} \otimes I + I \otimes \bm{b} \cdot \bm{\sigma}  + \sum_{i,j} T_{ij} \sigma_i \otimes \sigma_j \right),
	\label{eq:noncanonicalstate}
\end{equation}
where $\bm{\sigma}$ is the vector of Pauli matrices, $\bm{a}:=\langle \bm{\sigma} \otimes I \rangle$ is Alice's Bloch vector, $\bm{b}:= \langle I \otimes \bm{\sigma} \rangle$ is Bob's Bloch vector and $T :=\langle \bm{\sigma} \otimes \bm{\sigma}^T \rangle$ is the matrix of correlations, $i,j=1,2,3$.
The problem of state steerability reduces to the possible values of $\bm{a}, \bm{b}$ and $T$  for which there exists a LHS model for all possible measurements by Alice.
To simplify the problem, following Refs. \cite{Qui15,Bow16} we observe that Bob can apply a filtering operator $F := \rho_B^{-1/2} $ to his local state, resulting in the state
\begin{equation}
	\frac{ (I\otimes F) \rho (I\otimes F^\dagger)}{\Tr \left[ (I\otimes F) \rho (I\otimes F) \right]}, \label{eq:filter}
\end{equation}
which has a maximally mixed marginal on his side.
If $\rho_B$ is mixed, this operation is invertible and therefore must preserve steerability.
If $\rho_B$ is pure then clearly the state is non-steerable.
That is, we can take $\bm{b}=\bm{0}$ without loss of generality.
Furthermore, by allowing Alice to apply a local unitary on her qubit (which also preserves steerability), the correlation matrix can be taken to be diagonal $T=\text{diag}[t_1,t_2,t_3]$ without loss of generality.
Therefore, the so-called {\em canonical-state} \cite{Bow16} in the form
\begin{equation}
	\rho = \frac{1}{4}\left( I \otimes I + \bm{a} \cdot \bm{\sigma} \otimes I + \sum_{i} t_i \sigma_i \otimes \sigma_i \right),
	\label{eq:canonicalstate}
\end{equation}
captures all steerability properties of the shared state.
Thus, we see the question of determining two-qubit steerability amounts to specifying which values of $\bm{a}$ and $T$ admit a LHS model for Bob.
This remains a difficult open problem in general.

\subsection{Loss-Depleted States}
\label{lost}

By definition, a lossy channel maps every one-qubit state to the vacuum state $\ket{v}$ with some non-zero probability.
An initial state $\rho$ at the input of such a channel on Alice's side becomes the {\em loss-depleted} state
\begin{equation}
	\rho_{L} := \epsilon \rho + (1-\epsilon) \ketbra{v}{v} \otimes \rho_B. \label{eq:loststate}
\end{equation}
The quantity $\epsilon \in [0,1]$ is known as Alice's heralding efficiency, since it is the probability that Alice \emph{heralds} Bob's system for a measurement. 
Note that this quantity also subsumes information about the efficiency of Alice's detectors.
Its importance has been studied extensively in deriving loss-tolerant steering inequalities (see e.g. \cite{Eva13,Ben12}).
In a steering test, only loss on Alice's side is important, since Bob needs to consider only instances where he detects his system.

Henceforth, we restrict our discussion of $\rho$ to be a two-qubit state in the canonical form \eqref{eq:canonicalstate}.
Since Bob does not trust Alice in a steering test, he cannot acknowledge any claims Alice makes in regards to losing her qubit.
However, in order to give an honest Alice the best opportunity to demonstrate steering, it has been shown \cite{Ben12} that Bob should allow her to announce a {\em null result} `$0$', in addition to $\pm 1$.
That is, $X \in \{ -1, 0 ,1 \}$.
Clearly, an honest Alice who does not receive a qubit should always announce $X=0$, giving $\sigma_{0 | \bm{x}} = (1-\epsilon) I/2$.
If she receives her qubit, she performs the measurement requested by Bob.
Once such a measurement is performed and its outcome announced, Bob's assemblage in the lossy scenario now contains states, with $X \in \{ -1,1 \}$,
\begin{equation}
	\sigma_{X | \hat{\bm{x}}} = 
	\epsilon \Tr_A \left[ \left( E_{X|\bm{x}} \otimes I \right) \rho \right]. \label{eq:3assemblage}
\end{equation}
Hence, the pertinent question is: given a set of possible measurements, which loss-depleted states \eqref{eq:loststate} can Alice simulate by purely local means, given she can announce three outcomes?

\subsection{Connection to 1WS}
A key property of 1WS is that it requires some aspect of asymmetry in the shared state.
With the loss-depleted two-qubit scenario in mind, Evans and Wiseman \cite{Eva13} made an elegant link between the two-qubit Werner state \cite{Wer89} and one-way steering.
The two-qubit Werner state $\rho_W$ is defined as the mixture of the maximally mixed state with the singlet state $\ket{\Psi}:=2^{-1/2} (\ket{01}-\ket{10})$ by 
\begin{equation}
	\rho_W = \mu \ketbra{\Psi}{\Psi} + (1-\mu)\frac{I}{4},
\end{equation}
where $\mu \in [0,1]$.
The Werner state is steerable for $\mu > 1/2$ \cite{Wis07}.
Considering its loss-depleted counterpart, Evans and Wiseman \cite{Eva13} observed that Alice cannot steer Bob's state, with arbitrary projective measurements, if
\begin{equation}
	\epsilon \leq 2(1-\mu). \label{prxbound}
\end{equation}
However, if the roles of the parties are reversed, Alice simply considers her qubit sub-space in which she can be steered for $\mu > 1/2$---that is, passing Alice's qubit through any sufficiently lossy channel makes it one-way steerable.
Here, we generalise the fundamental idea of Evans and Wiseman to general two-qubit states by explicitly constructing LHS models for states of the form
\begin{equation}
\begin{aligned}
	\rho_{L} &:= \frac{\epsilon}{4}\left( I + \bm{a} \cdot \bm{\sigma} \otimes I + \sum\limits_{i=x,y,z} t_i \sigma_i \otimes \sigma_i \right) \\
	&+ (1-\epsilon) \ketbra{v}{v} \otimes \frac{I}{2}	\label{eq:lds}
\end{aligned}
\end{equation}
based on the method of Bowles {\em et. al.} \cite{Bow16}.

\section{Local-Hidden-State Models for Loss-Depleted States\label{lhsmodel}}

Before presenting our main result, we observe the following.
\def \lmax{\Lambda_{\text{max}}}
\def \lmin{\Lambda_{\text{min}}}
\begin{lemma}
Let $\lmax(\sigma_{X| \bm{x}})$ denote the maximum eigenvalue of the two-qubit state $\sigma_{X | \bm{x}}$ which is a member of the assemblage $\{ \{ \sigma_{X | \bm{x}} : X \} : \bm{x} \}$, for which $\sigma_{+1| \bm{x}} + \sigma_{-1| \bm{x}} = \epsilon I/2 \quad\forall \bm{x}.$
Then, the entire assemblage of such states can be reproduced by a LHS model for Bob if and only if there exists an assemblage $\{ \{ \sigma^\text{LHS}_{X | \bm{x}} : X \} : \bm{x} \}$, with $\Tr\left[ \sigma^{\text{LHS}}_{+1 | \bm{x}} \right] = \Tr\left[ \sigma_{+1 | \bm{x}} \right] \forall \bm{x}$ such that 
\begin{equation}
	\lmax\left(\sigma^{\text{LHS}}_{+1 | \bm{x}}\right) \geq \lmax\left(\sigma_{+1 | \bm{x}} \right) \qquad \forall \bm{x}. \label{eq:eiglemma}
\end{equation}
\end{lemma}
\noindent A proof is given in the Appendix.

Next, we derive a simple constraint which, if satisfied, ensures that a loss-depleted two-qubit state \eqref{eq:lds} is non-steerable from Alice to Bob.
We allow Alice's possible measurements to be arbitrary projective measurements, but not arbitrary POVMs. 
\newtheorem{thm}{Theorem}
\begin{thm}
Consider the loss-depleted two-qubit state \eqref{eq:lds}.
If
\begin{equation}
	\max\limits_{\bm{\hat{x}} \in \widehat{\mathbb{R}^3}} \left[ (1-\epsilon) \left| \bm{a} \cdot \bm{\hat{x}} \right| + \frac{\epsilon}{2} \left( 1 + \left( \bm{a} \cdot \bm{\hat{x}} \right)^2 \right) +  \Vert T \bm{\hat{x}} \Vert \right] \leq 1, \label{eq:result}
\end{equation}
where $\widehat{\mathbb{R}^3}$ is the set of unit vectors in $\mathbb{R}^3$, $\rho_L$ is non-steerable from Alice to Bob considering arbitrary projective measurements.
\end{thm}
\newtheorem*{remark}{Remark}
\begin{remark}
For $\epsilon=1$, this condition reduces to the inequality in Ref. \cite{Bow16}.
\end{remark}
\begin{proof}
Our proof is similar to that of Bowles {\em et. al.} \cite{Bow16}, with one subtle difference.
We will proceed in two steps.
First, we calculate the steered states (in particular, their largest eigenvalues) prepared by a projective measurement by Alice.
Since each two-qubit projective measurement operator $\Pi_{X | \hat{\bm{x}}}$ can be represented by a unit vector $\hat{\bm{x}}$ on the Bloch sphere, we denote $\bm{x}$ by $\hat{\bm{x}}$.
Second, we will show that the steered states arising can be simulated by a LHS model for Bob.

\emph{Case 1: Alice's qubit is lost.}
First, we consider the case where Alice loses her qubit.
Here, when Bob announces a measurement $\bm{x}$ to perform, she should announce the null result with probability unity \cite{Eva13}, as discussed in \S \ref{lost}.
Bob's ``steered'' state will average to the random state, meaning his assemblage will contain the state $I/2$ with fraction $(1-\epsilon)$.
That is, $\sigma_{0 | \hat{\bm{x}}} = (1-\epsilon)\frac{I}{2}$.

\emph{Case 2: Alice receives a qubit.}
If Alice's qubit is not lost into the vacuum, she proceeds by performing a projective measurement requested by Bob.
The steered states for $X=+1$ are \cite{Jev15}
\begin{equation}
	\epsilon\Tr_A \left[ \left( \Pi_{X|\hat{\bm{x}}} \otimes I \right) \rho \right]= \frac{\epsilon}{4} \left[ (1 + \bm{a} \cdot \hat{\bm{x}}) I + T\hat{\bm{x}}\cdot \bm{\sigma} \right],
\end{equation}
which can be diagonalized \cite{Bow16} by rotating to a frame defined by the basis $\{\ket{\hat{s}}, \ket{-\hat{s}} \}$ with Bloch vector $\hat{\bm{s}} = T\hat{\bm{x}} / \Vert T\hat{\bm{x}} \Vert$.
Its eigenvalues are found to be
\begin{equation}
	\lambda_\pm = \frac{\epsilon}{4} \left( 1 + \bm{a} \cdot \hat{\bm{x}} \pm \Vert T\hat{\bm{x}} \Vert \right).
\end{equation}
Thus, 
\begin{equation}
	\lmax\left(\sigma_{+1 | \hat{\bm{x}}}\right) = \frac{\epsilon}{4} \left( 1 + \bm{a} \cdot \hat{\bm{x}} + \Vert T\hat{\bm{x}} \Vert \right). \label{eq:asseig}
\end{equation}

Next, we construct a LHS model for this assemblage.
Let the ensemble of LHSs be uniformly distributed on the Bloch sphere, and Alice's cheating strategy be as illustrated in Fig. \ref{fig:circle}.
Define the coordinate system $(z, \phi)$ such that the $\hat{z}$ axis is aligned along $\hat{\bm{s}}$.
If $\epsilon-1 \leq \hat{\lambda} \cdot \hat{\bm{s}} \leq 1-\epsilon$, she announces the null result. 
This ensures Case 1 above will be satisfied.

To simulate Case 2 above, her strategy in other regions of the sphere depends on the measurement request by Bob (see Fig. \ref{fig:circle}).
As above, we only need to describe the case $X=+1$.
Let this region of the Bloch sphere be denoted by $\mathcal{R}_+$.
The region $\mathcal{R}_+$ depends on $\delta(\hat{\bm{x}})$, illustrated in Fig. \ref{fig:circle}.
In particular, its value is constrained by the statistics $\Tr[ \Pi_{X|\hat{\bm{x}}} \rho_A]$ of Alice's reduced state $\rho_A$ that she must simulate.
That is,
\begin{equation}
	\frac{\epsilon}{2}\left( 1 + \bm{a} \cdot \bm{\hat{\bm{x}}} \right) \equiv \int_{\mathcal{R}_+} \frac{dz d\phi}{4\pi}. \label{eq:marginal}
\end{equation}
Evaluating the integral over the regions depicted in Fig. \ref{fig:circle}, it is straightforward to show
\begin{equation}
	\delta(\hat{\bm{x}})= 
	\begin{cases}
	-\epsilon \bm{a} \cdot \bm{\hat{\bm{x}}} & \text{if} \quad p(+|\hat{\bm{x}},\lambda) \leq p(-|\hat{\bm{x}},\lambda),  \\
         \phantom{-}\epsilon \bm{a} \cdot \bm{\hat{\bm{x}}} & \text{if} \quad p(+|\hat{\bm{x}},\lambda) > p(-|\hat{\bm{x}},\lambda),
	\end{cases}
	\label{eq:deltas}
\end{equation}
where negative values of $\delta(\hat{\bm{x}})$ should be understood as reductions in size of the region for which $X=+1$.
\begin{figure}[htbp]
\begin{center}
	\includegraphics[width=.8\linewidth]{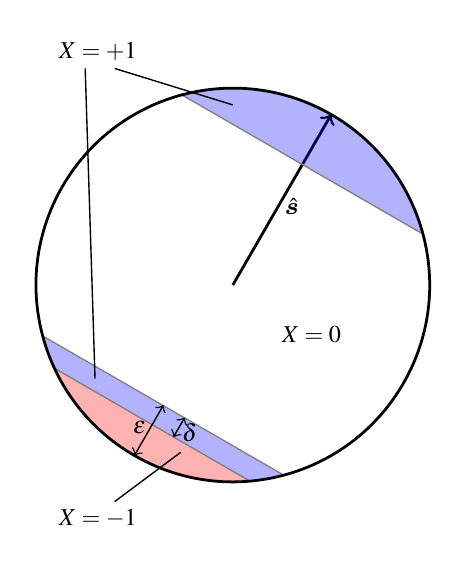}
\caption{Alice's cheating strategy in our LHS model.
Given a measurement from Bob corresponding to some unit vector $\bm{x}$ on the Bloch sphere, Alice announces either $+1, 0$ or $-1$ depending on the containing region of the LHS $\hat{\lambda}$ (not depicted).
The size of the polar caps vary by an amount $\delta(\hat{\bm{x}})$, allowing her to simulate a non-maximally mixed marginal.}
\label{fig:circle}
\end{center}
\end{figure}

We represent the LHS as a pure state in cylindrical coordinates $(z,\phi)$ within the steered-state basis $\ket{\pm \hat{s}}$,
\begin{equation}
	\ket{\hat{\lambda}(z,\phi)} = \sqrt{\frac{1+z}{2}}\ket{\hat{s}} + \sqrt{\frac{1-z}{2}} e^{i\phi} \ket{-\hat{s}}.
\end{equation}
Then, the predicted assemblage (for $X=+1$) is
\begin{equation}
	\sigma_{+1 | \hat{\bm{x}}}^{\text{LHS}} = \int\limits_{\mathcal{R}_{+}}  \frac{dz d\phi}{4\pi}  \ketbra{\hat{\lambda}(z,\phi)}{\hat{\lambda}(z,\phi)}. \label{eq:predass}
\end{equation}
Since $\int_0^{2\pi} d\phi e^{i\phi}=0$ the off-diagonal elements vanish, and therefore the largest eigenvalue of \eqref{eq:predass} is simply
\begin{align}
	\Lambda_\text{max} \left( \sigma_{+ | \hat{\bm{x}}}^{\text{LHS}} \right) &= \frac{1}{2} \int\limits_{\mathcal{R}_+} dz \frac{1+z}{2} \\
	&= \frac{\epsilon}{4}\left(\left( 1 + \bm{a} \cdot \bm{\hat{\bm{x}}} \right)+  \frac{1}{\epsilon}\int\limits_{\mathcal{R}_+} dz z \right), \label{eq:midway}
\end{align}
where we have substituted Eq. \eqref{eq:marginal}.
Comparing \eqref{eq:asseig} and \eqref{eq:midway} in light of Lemma 1, $\lmax\left(\sigma^{\text{LHS}}_{+1 | \hat{\bm{x}}}\right) \geq \lmax\left(\sigma_{+1 | \hat{\bm{x}}}\right)$ requires that
\begin{equation}
	\Vert T\hat{\bm{x}} \Vert \leq \frac{1}{\epsilon}\int\limits_{\mathcal{R}_+} dz z. \label{eq:anti}
\end{equation}

Due to the anti-symmetry of the integrand in \eqref{eq:anti}, the two cases outlined in \eqref{eq:deltas} integrate to the same value. 
This allows us to write $\int_{\mathcal{R}_+} dz z   = \int_{1-\epsilon-|\delta(\hat{\bm{x}})|}^1 dzz.$
Upon evaluating this integral, inequality \eqref{eq:anti} then implies
\begin{equation}
	\Vert T\hat{\bm{x}} \Vert + (1-\epsilon)|\bm{a}\cdot \hat{\bm{x}}| +  \frac{\epsilon}{2}\left( 1 + \left( \bm{a} \cdot \hat{\bm{x}}^2 \right) \right)\leq 1,
\end{equation}
which can be ensured for all measurements by maximizing the left-hand-side over $\hat{\bm{x}}$.
Thus, \eqref{eq:result} follows.
\end{proof}

In the following section, we will extend our LHS for arbitrary projective measurements to one for the most general measurements, and apply our result to a recent one-way steering experiment.

\section{Extension to POVMs and an Experiment. \label{povms}}

Our sufficient criterion \eqref{eq:result} ensures Alice cannot steer Bob for arbitrary projective measurements.
Here, we extend our condition to ensure non-steerability of loss-depleted two-qubit states under POVMs.
In order to make this extension, we can exploit Lemma 1 derived in Ref. \cite{Qui15}:

\begin{lemma} \label{povmlemma}
Let $\rho$ be a quantum state acting on $\mathcal{H}^d \otimes \mathcal{H}^d$ for which there exists a LHS model for projective measurements, from Alice to Bob.
Then, the state
\begin{equation}
	\rho' = \frac{1}{d+1} \left( \rho + d\pi_\perp \otimes \rho_B  \right) \in \mathfrak{B} \left(\mathcal{H}^{d+1} \otimes \mathcal{H}^d \right),
\end{equation}
admits a LHS model allowing for POVMs from Alice to Bob, if $\pi_\perp$ is a projector onto a subspace orthogonal to the support of $\rho_A$.
\end{lemma}
We now extend \eqref{eq:result} to ensure non-steerability of $\rho_L$ for POVMs.
To this end, we can consider the loss-depleted two-qubit state to act on $\mathcal{H}^3 \otimes \mathcal{H}^3$, where $\Tr_A \left[ \rho_L \right]$ is supported only on $\mathcal{H}^2$.
Identifying $\pi_\perp$ as the projector onto another vacuum state and consequently denoting $\rho'$ as $\rho_L'$, it follows from Lemma \ref{povmlemma} that the state
\begin{equation}
	\rho'_L = \frac{\epsilon}{4}\rho + \left( 1 - \frac{\epsilon}{4} \right) \ketbra{v}{v} \otimes \frac{I}{2}
\end{equation}
is non-steerable for POVMs, if $\rho_L$ is non-steerable for projective measurements.
(Note that this equation is the correct expression for Eq. (5) in Ref. \cite{Wol16}, where factor of 1/3 instead of 1/4 was used.
Or, at least, it was not shown in Ref. \cite{Wol16} that a factor of $1/3$ is sufficient.)
In other words, reducing Alice's heralding efficiency by a factor of $1/4$ makes any state \eqref{eq:lds} satisfying inequality \eqref{eq:result} non-steerable from Alice to Bob for arbitrary POVMs, {\em i.e.} if 
\begin{equation}
	\max\limits_{\bm{\hat{x}} \in \widehat{\mathbb{R}^3}} \left[ \left(1-4\epsilon\right) \bm{a} \cdot \hat{\bm{x}} + 2 \epsilon \left( 1 + \left( \bm{a} \cdot \hat{\bm{x}} \right)^2 \right) +  \Vert T \hat{\bm{x}} \Vert \right] \leq 1 \label{eq:povmresult}
\end{equation}
is satisfied.

\subsection{Analysis of experimental data}

In Ref. \cite{Wol16}, the first one-way steering experiment involving arbitrary projective measurements and POVMs was reported.
Note that, in this work we have chosen the convention of Alice being unable to steer Bob, whilst this experiment aimed to establish that Bob could not steer Alice, while Alice did steer Bob.
Thus we swap the roles of the experiment's Alice and Bob in applying Eqs. \eqref{eq:result} and \eqref{eq:povmresult} to the states engineered in the experiment.

Two demonstrations of one-way EPR-steering were reported to which our results are relevant---the first where the steering party (our Alice) was permitted to make arbitrary projective measurements, and the second where she can make POVM measurements.
In the experimental setup, pairs of entangled photons were generated and distributed between Alice and Bob.
These states were engineered and found by tomographical reconstruction to have a high fidelity with a Werner state---$(99.6\pm 0.1)\%$ and $(99.1\pm 0.3)\%$ in the projective measurement and POVM tests, respectively.
In addition to the expected photon losses on both sides, Alice's qubit was passed through a high-loss channel, resulting in overall heralding efficiencies of $\epsilon = 0.022 \pm 0.006$ (projective measurements) and $\epsilon = 0.005 \pm 0.003$ (POVMs).
Originally, these heralding efficiencies, together with the parameters of the nearest Werner state were argued to imply non-steerability \cite{Wol16}.
Here, we seek to apply our LHS models, which make no assumptions on the two-qubit state, to test this conclusion.

To calculate the relevant quantity in Eqs. \eqref{eq:result} and \eqref{eq:povmresult}, with uncertainties, we follow the method of \cite{Wol16}.
That is, we simulate the Poissonian errors in the data by using the tomographically reconstructed state to generate an ensemble of 200 states consistent with the data.
For our calculations, we take the ``worst case'' (highest) value of $\epsilon$, given by the best estimate plus the experimental error, since reducing the efficiency of transmission can only reduce Alice's ability to steer Bob.
For each state in the ensemble, we first calculate $\bm{a}, T$ and $\bm{b}$.
Then, after performing the filtering operation on Bob's side given by \eqref{eq:filter}, we diagonalise the correlation matrix $T$.
Finally, we numerically perform the maximization over $\hat{\bm{x}}$ in either \eqref{eq:result} or \eqref{eq:povmresult}, depending on the case.
To compare the data to our condition under a quantified error, we take the list of all such maximizations and calculate its mean and standard deviation.

Upon evaluating the left-hand-side for all iterations, we find {\em violation} of our condition for the experimental data in both the projective measurement and POVM cases.
That is, the conditions we have derived are not strong enough to prove non-steerability of the experimental data in either case.
However, inserting the experimental data from the POVM experiment into our condition for non-steerability under projective measurements \eqref{eq:result}, we calculate the inequality to require $(0.9997 \pm 0.0067) \leq 1$.
Thus, the data in this case satisfies our condition, but not with any significance.

One might conjecture that the data could have satisfied our condition simply by decreasing $\epsilon$ to an arbitrarily low, non-zero value.
A natural question, then, is to ask: given a {\em steerable} two-qubit state, is it always possible to pass Alice's qubit through a lossy channel with a non-zero value of $\epsilon$ such that the resulting loss-depleted state is non-steerable by \eqref{eq:result}?
That is, does there always exist a heralding efficiency $\epsilon_0 >0$, given by
\begin{equation}
	\epsilon_0 = \frac{1 - \Vert T\bm{\hat{x}_0} \Vert - |\bm{a}\cdot \bm{\hat{x}_0}|}{(1+\left( \bm{a} \cdot \bm{\hat{x}_0} \right)^2)/2 - |\bm{a}\cdot \bm{\hat{x}_0}|}
	\label{eq:e_0}
\end{equation}
for all $T, \bm{a}, \bm{\hat{x}_0}$, where $\bm{\hat{x}_0}$ is the unit vector which maximizes \eqref{eq:result}.
Notice that the denominator of \eqref{eq:e_0} is strictly greater than zero, except when $\bm{a} \cdot \bm{\hat{x}_0}=\pm1$.
However, if this is true then $T=0$ by the non-negativity of \eqref{eq:lds} and therefore the state must always be non-steerable.
When $\bm{a} \cdot \bm{\hat{x}_0} \neq \pm1$, the question reduces to whether the numerator is always positive.
To test this, we performed a search over the two-qubit state space to see if $\Vert T\bm{\hat{x}_0} \Vert + |\bm{a}\cdot \bm{\hat{x}_0}| <1$ for all $T, \bm{a}, \bm{\hat{x}_0}$.
Interestingly, we found parameters for which this does not hold---hence, the answer to this conjecture is no.
Therefore, it would be reasonable to conclude that rigorously proving non-steerability from the data reported in Ref. \cite{Wol16} would be impossible, even if arbitrary losses were available.
In any case, if it were possible, it would require the construction of a better LHS model for this class of states than we have assumed.

\section{Conclusion \label{conclusion}}

We have provided criteria which are sufficient to prove a two-qubit state is non-steerable if the steering party suffers from loss.
Although our construction ensures non-steerablity under arbitrary projective measurements, we have extended it to allow for POVMs.
Since loss is inevitable in any one-way steering experiment, our main results should find practical use.
They are rigorous, in the sense that they allow a physicist to go beyond fidelity measures to prove non-steerability.
However, in terms of the experimental data in Ref. \cite{Wol16}, proving non-steerability remains an open question.

As a future direction, it would be interesting to develop complementary loss-tolerant steering inequalities allowing for non-maximally-mixed marginals on Bob's side.
Together with our result, these would uncover states which exhibit 1WS in a robust manner---that is, with the ability to account for {\em any} variations in the state which unexpectedly arise from experimental imperfections.
Another natural extension to this work would be the construction of a LHS model for loss-depleted states that relaxes the assumption of a uniform distribution of LHSs on Bob's Bloch sphere.

\appendix*
\label{lemma1proof}
\section{Proof of Lemma 1}
We will prove Lemma 1 by showing Alice can match the assemblage $\{ \{ \sigma_{X | \bm{x}} : X \} : \bm{x} \}$ from the assemblage $\{ \{ \sigma^\text{LHS}_{X | \bm{x}} : X \} : \bm{x} \}$ by altering her response function $p^{\text{LHS}}(X|\bm{x},\lambda)$, assuming Eq. \eqref{eq:eiglemma} holds.
To this end, we allow her to ``flip'' the outcome she announces, contrary to her cheating strategy in the LHS model.
We introduce flipping probabilities $f$ and $g$, which describe the likelihood she flips between announcing $+1$ and $-1$, respectively.
Note that only the states in the steered assemblage for $X=+1$ need to be reproduced, since $X=-1$ will be satisfied by $\sigma_{-1| \bm{x}}= \epsilon I/2-\sigma_{+1| \bm{x}}$.
For simplicity, we shall write $p(+1|\bm{x})$ as $p_+$.
Expressing the states in the assemblage in a basis in which it is diagonal for a given measurement direction, the resulting portion of the assemblage for $X=+1$ is then
\begin{equation}
\begin{aligned}
	p_{+}\left(
\begin{array}{cc}
 \alpha_+ & 0  \\
 0 & 1-\alpha_+ 
\end{array}
\right) &= fp_{+}\left(
\begin{array}{cc}
 \alpha_+^{\text{LHS}} & 0  \\
 0 & 1- \alpha_+^{\text{LHS}}
\end{array}
\right) \\
&+  g(1-p_{+})\left(
\begin{array}{cc}
 \alpha_-^{\text{LHS}} & 0  \\
 0 & 1-\alpha_-^{\text{LHS}}  
\end{array}
\right).
\label{eq:fg}
\end{aligned}
\end{equation}
Here, $p_+ \alpha_+^{\text{LHS}}$(resp. $p_- \alpha_-^{\text{LHS}}$) denotes an eigenvalue of the LHSs $\sigma^\text{LHS}_{+1}$ ($\sigma^\text{LHS}_{-1}$) and $p_+ \alpha_+$ is an eigenvalue of $\sigma_{+1}$.
Without loss of generality, we assume $p_+ \alpha_+^{\text{LHS}}$ and $p_+ \alpha_+$ are largest eigenvalues of their respective states.
Note we have dropped the $\bm{x}$-dependence, since relations similar to \eqref{eq:fg} exist for each $\bm{x}$.
Without flipping probabilities, the condition $\lmax\left(\sigma^{\text{LHS}}_{+1 | \bm{x}}\right) \geq \lmax\left(\sigma_{+1 | \bm{x}} \right)$ translates to $\alpha_+ < \alpha_+^{\text{LHS}}$.
Assuming this, we wish to show that there will always exist $f,g\in[0,1]$ such that \eqref{eq:fg} can be satisfied. 

Since $\sigma^{\text{LHS}}_{+1| \bm{x}} + \sigma^{\text{LHS}}_{-1| \bm{x}} = I/2$, we have
\begin{equation}
	\alpha_-^{\text{LHS}}=\frac{1/2 -p_+\alpha_+^{\text{LHS}}}{1-p_+},
\end{equation}
and by tracing over both sides in \eqref{eq:fg}
\begin{equation}
	g=\frac{p_+(1 - f)}{1-p_+}. \label{eq:g}
\end{equation}
Substituting these quantities into \eqref{eq:fg}, we solving for the flipping probability
\begin{equation}
	f= \frac{\alpha_+ + C}{\alpha_+^{\text{LHS}} + C},
\end{equation}
where $C:=(p_+\alpha_+^{\text{LHS}} - 1/2)/(1-p_+)$.
Since $p_+\alpha_+^{\text{LHS}} > 1/2$ and $\alpha_+ < \alpha_+^{\text{LHS}}$ by assumption, $f\in[0,1]$.
In turn, \eqref{eq:g} then implies $g\in[0,1]$ so that \eqref{eq:fg} can always be satisfied by choosing appropriate flipping probabilities, for each $\bm{\hat{x}}$.

Furthermore, the conditions stipulated in Lemma 1 must also be necessary for steerability, since, if there exists an assemblage for which there does not exist {\em any} LHS model such that Eq. \eqref{eq:eiglemma} is satisfied, it follows (by the definition of steerability) that the corresponding two-qubit state must be steerable.

\begin{acknowledgements}
T. J. B. and S. W. acknowledge support by an Australian Government Research Training Program (RTP) Scholarship.
This work was supported by the ARC Centre of Excellence CE110001027.
\end{acknowledgements}

\bibliographystyle{apsrev4-1}
\bibliography{lossylhs.bib}


\end{document}